\documentclass[aip,reprint,jap]{revtex4-1}

\usepackage{graphics}
\usepackage{soul}

\begin{document}

\title{
Relevance of $4f$-$3d$ exchange
to finite-temperature magnetism
of rare-earth permanent magnets:
an {\it ab-initio}-based spin model
approach 
for
NdFe$_{12}$N}

\author{Munehisa~Matsumoto$^{1}$,
Hisazumi~Akai$^{1,2}$, 
Yosuke~Harashima$^{1,3}$, Shotaro~Doi$^{1,2}$,
Takashi~Miyake$^{1,3}$
}
\affiliation{$^1$Elements Strategy Initiative Center for Magnetic Materials (ESICMM),
National Institute for Materials Science (NIMS),
Sengen 1-2-1, Tsukuba 305-0047, JAPAN\\
$^{2}$ Institute for Solid State Physics, University of Tokyo, Kashiwa 277-8581, JAPAN,\\
$^{3}$ National Institute of Advanced Industrial Science and Technology (AIST),
Umezono 1-1-1, Tsukuba 305-8568, JAPAN\\
}

\date{\today}

\begin{abstract}
A classical spin model derived {\it ab initio} for 
rare-earth-based permanent magnet compounds is presented.
Our target compound, NdFe$_{12}$N,
is
a material
that goes beyond today's champion magnet compound Nd$_{2}$Fe$_{14}$B
in its intrinsic magnetic properties with a simpler crystal structure.
Calculated temperature dependence of the magnetization and the anisotropy field
agree with the latest experimental results in the leading order.
Having put the realistic observables under our numerical control,
we propose that
engineering $5d$-electron-mediated indirect exchange coupling between $4f$-electrons in Nd and $3d$-electrons from Fe
would most critically
help to enhance the material's utility over the operation-temperature range.
\end{abstract}


% 75.50.Ww Permanent magnets
% 75.30.Gw Magnetic anisotropy
%
% 75.10.Lp Band and itinerant models
% 75.10.Hk classical spin models
%
% 71.15.Rf Relativistic effects

\maketitle

\section{Motivations}
In the past three decades,
Nd$_{2}$Fe$_{14}$B~\cite{sagawa_1984_sintering}
has been the champion permanent magnet compound.
A drawback of Nd$_{2}$Fe$_{14}$B has been its relatively low Curie temperature
and that
some practical applications require replacements of Nd with heavy-rare-earth (HRE) elements
such as Dy and Tb to enhance the high-temperature
coercivity, which is roughly proportional
to the anisotropy field~\cite{hirosawa_1986}.
Since the HRE elements are less abundant,
ways to achieve the equivalent magnetic properties to those of HRE-doped permanent magnets
using only light rare-earth (LRE) elements have recently been sought after~\cite{hono_2012}. Also the relevance
of understanding and controlling
the finite-temperature magnetism of $4f$-$3d$ intermetallics is appreciated ever more.

A possible solution was recently
suggested~\cite{miyake_2014}
by the material NdFe$_{12}$N stabilized in an almost
bulk state~\cite{hirayama_2014,hirayama_2015},
where a newly fabricated film sample
that consists of more than hundred unit-cell layers
shows superior intrinsic magnetization and anisotropy field to Nd$_{2}$Fe$_{14}$B.
The materials family, RFe$_{12-x}$T$_{x}$(N) (R=rare earth),
had actually been known for a long time~\cite{1-12_review}
where the nitrogenation
pulls up the Curie temperature by 100-200~[K]
and the magnetic anisotropy is enhanced as well,
but the achieved magnetic properties were not on a par with the champion magnet compound
Nd$_2$Fe$_{14}$B partially
due to the necessity for the presence of the third element T=Ti etc.
to stabilize the particular crystal structure.
Recent breakthrough~\cite{hirayama_2014} made it possible to have 
NdFe$_{12}$ 
without the third element in a sample fabricated as a thick film and nitriding achieved the intrinsic
magnetic properties that goes beyond Nd$_2$Fe$_{14}$B at high temperatures~\cite{hirayama_2014}.

Thus we are motivated to 
theoretically address the finite-temperature magnetism 
of NdFe$_{12}$N. This would provide the prospect for its intrinsic
magnetic properties, which serves to solve the high-temperature coercivity problem
in LRE-based compounds.
On the theory side,
finite-temperature magnetism of
rare-earth-based
permanent-magnet materials poses
a fundamentally challenging many-body problem:
{\it ab initio} predictions
mostly focus on the ground-state properties
and the finite-temperature magnetism was discussed
at best on the basis of a mean-field theory
of a simplified model on the basis of
a molecular field acting on an isolated rare-earth
magnetic moment~\cite{yamada,cadogan_1988,radwanski,kuzmin,faehnle,sasaki_2015}.
For comparison with experimentally observed magnetic anisotropy, contributions from 3$d$ electrons are 
sometimes added in an {\it ad hoc} manner.
In principle, the theory of magnetism in $4f$-$3d$ intermetallics
takes a description of the correlated electrons
in $4f$ and $3d$-orbitals, which may be done with
a multi-orbital
periodic Anderson model (PAM) with the conduction electrons composed of
$5d$-band out of the rare-earth elements, harboring
two species of impurities, $3d$ and $4f$, each with the different levels
of on-site electronic correlation.

In order to meet the urgent practical needs
and also to provide a guideline data for future realistic simulations of
PAM with huge number of orbitals,
we exploit
the essence of
a simplified model~\cite{skomski_1998}
to describe only the low-energy effective physics
of $4f$-$3d$ intermetallics
with the model parameters determined as realistically as possible through
{\it ab initio} calculations:
we define
a multi-sublattice spin model
with one group of the sublattices describing the $4f$-originated
localized magnetic moments
and the other describing the $3d$-magnetization;
the spins reside on a realistic lattice
that mimicks the crystal structure of the given target material NdFe$_{12}$N
with the {\it ab initio} input parameters.
Note that the
effective parameters, such as the strength of exchange couplings and
crystalline electric fields, are determined as a consequence of
the electronic states. They are target-material dependent,
thus first-principles evaluation of the parameters
is crucial for the quantitative modeling.
Then we solve
the finite-temperature many-body problem
with numerically exact Monte Carlo method
to get the temperature dependence
of magnetic observables
and quantitatively compare with the latest experimental data.
Our realistic lattice model
is more realistic than were discussed in the previous works:
The sublattice-specific character of each of the Nd and Fe atoms
in the unit cell is taken into account on the basis of first principles,
which is in contrast to a uniform molecular field imposed by Fe
acting on rare-earth magnetic moments.
Establishing the computational
control of the intrinsic properties of magnetism of NdFe$_{12}$N,
we discuss within our model
how to manipulate it to enhance its practical utility
most effectively.

\section{The realistic-lattice spin model} 
The spin model Hamiltonian defined on the lattice
of the given crystal structure
reads as follows.
\begin{eqnarray}
{\cal H} &  = & {\cal H}_{\rm T} + {\cal H}_{\rm R} + {\cal H}_{\rm RT},
\label{eq::realistic_two_sublattice_model}\\
{\cal H}_{\rm T} &  = &
  -\sum_{\left<i,j\right>\in {\rm T}}\left(2J_{ij}^{\rm TT}\right){\mathbf S}_{i}\cdot{\mathbf S}_{j}
-\sum_{i\in {\rm T}}D_{i}^{\rm T}\left(S_{i}^{z}\right)^2, \label{eq::3d}\\
{\cal H}_{\rm R} &  = & -\sum_{m\in {\rm R}}D_{m}^{\rm R}\left(J_{m}^{z}\right)^2,\label{eq::4f}\\
{\cal H}_{\rm RT} &  = & \alpha_{\rm RT}
  \sum_{\left<m,i\right>,m\in{\rm R},i\in {\rm T}}\left(2J_{mi}^{\rm RT}\right)
(g_{J}-1) {\mathbf J}_{m}
\cdot
{\mathbf S}_{i}
\label{eq::J_RT}
\end{eqnarray}
We have denoted the lattice points on which $3d$ ($4f$) magnetic moments reside by T (R),
respectively. 
Here ${\mathbf S}_{i}$ is a magnetic moment defined on site $i$,
$J_{ij}$ is the exchange coupling between localized
magnetic moments ${\mathbf S}_i$ and ${\mathbf S}_j$, and 
$D_{i}$
encodes the single-ion
magnetic anisotropy energy (MAE).
Note that
the summation $\sum_{\left<i,j\right>}$ runs
over each bond connecting the sites $i$ and $j$
only once.
The $4f$-part is described explicitly
with the total magnetic moment ${\mathbf J}={\mathbf L}+{\mathbf S}$
with ${\mathbf L}$ being
the orbital moment.
The spin moment can be extracted via ${\mathbf S}=(g_{J}-1){\mathbf J}$
with $g_{J}$ being Land\'{e}'s $g$-factor.
Describing a $4f$-moment of Nd$^{3+}$
with a classical spin of length $g_{J}\sqrt{J(J+1)}$ with $J=9/2$
is semi-quantitatively
justified within the scope of setting the target temperature range
in $200~\mbox{[K]}\stackrel{<}{\sim}T\stackrel{<}{\sim}500~\mbox{[K]}$,
with the $J$-multiplets separated in the scale of $1000~\mbox{[K]}$.
The $4f$-$4f$ exchange coupling terms, $J^{\rm RR}$,
in the scale of $O(1)$~[K]
have been dropped because our target energy scale
to be realistically described is motivated by the typical operating
temperature range $200~\mbox{[K]} \stackrel{<}{\sim} T \stackrel{<}{\sim} 500~\mbox{[K]}$
of permanent magnets. Magnetic energy scales coming from
$J^{\rm RR}$'s are at most $\sim 0.1 T$ and should be washed out.

The $4f$-$3d$ indirect
exchange coupling as denoted by $J_{mi}^{\rm RT}$ in Eq.~(\ref{eq::J_RT})
comes from $(5d)^{m}$-$(3d)^n$ ($m\stackrel{<}{\sim}1$ and $n>5$)
exchange which is antiferromagnetic and RE on-site $4f$-$5d$ direct exchange which is ferromagnetic.
Overall $J_{mi}^{\rm RT}$ is an antiferromagnetic coupling between the spin component of $4f$
and $3d$, which means $4f$ total moment and $3d$ magnetization are
ferromagnetically coupled for Nd$^{3+}$ with $g_{J}=8/11$.
The overall scale factor $\alpha_{\rm RT}$,
tentatively set to be one,
has been introduced to
phenomenologically describe
the indirect nature of the $4f$-$3d$ exchange coupling.
% in a phenomenological way.

\begin{figure}
\begin{center}
\scalebox{0.4}{
\includegraphics{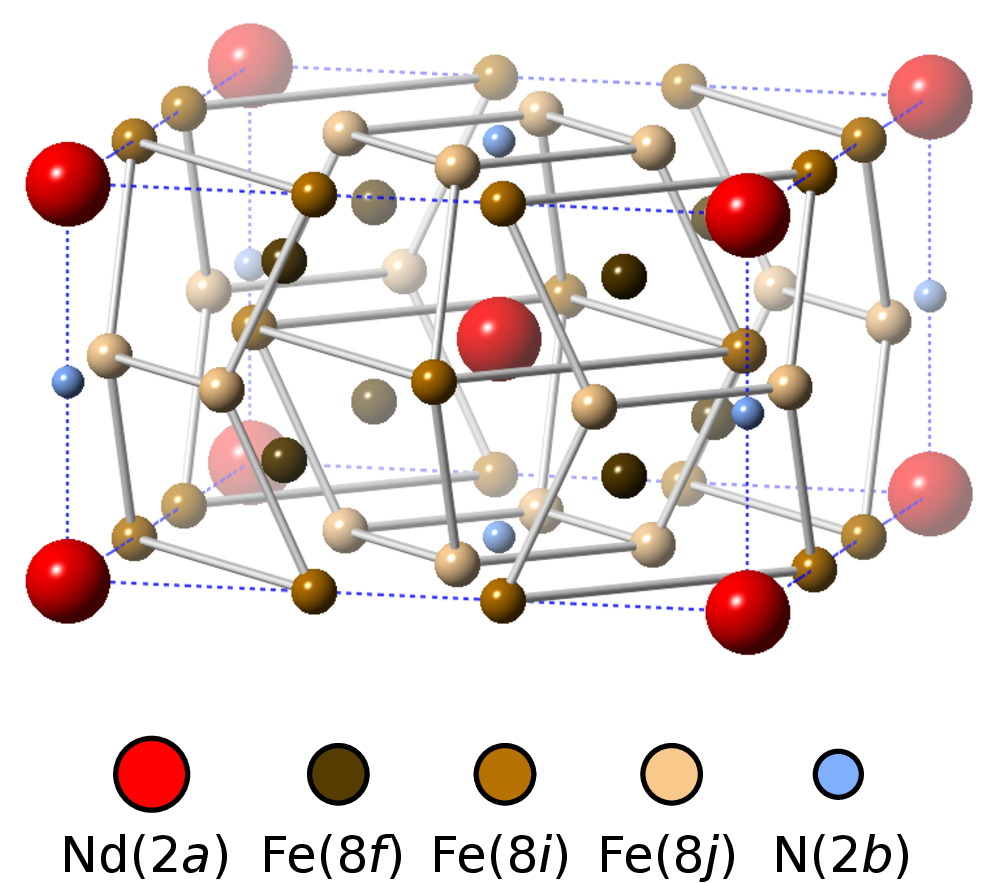}}\\
\end{center}
\caption{\label{fig::Harashima_lattice} Crystal structure of NdFe$_{12}$N. Along
the Nd($2a$)-N($2b$)-Nd($2a$) direction runs the $c$-axis. Perpendicular to it,
the other two equivalent directions, $a$-axis and $b$-axis,
span the Nd($2a$)-Fe($8i$)-Fe($8i$)-Nd($2a$) lines.
}
\end{figure}
\begin{figure}
\begin{center}
\begin{tabular}{l}
(a)\\
\scalebox{0.7}{
\includegraphics{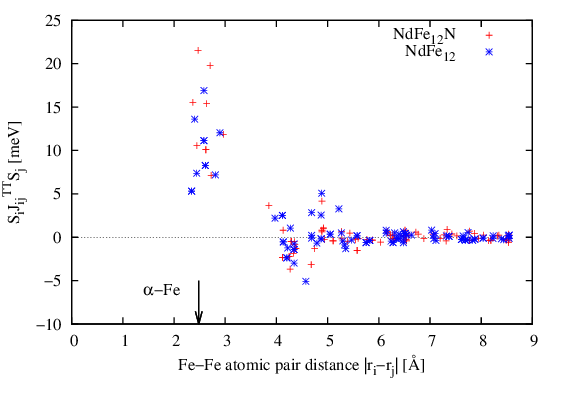}}\\
(b) \\
\scalebox{0.7}{
\includegraphics{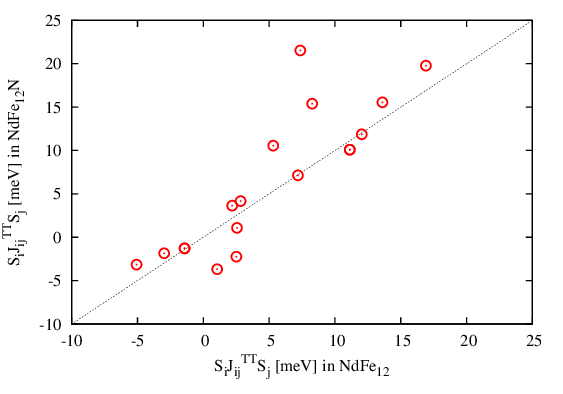}}
\end{tabular}
\end{center}
\caption{\label{fig::Akai_Jij} (a) Calculated Fe-Fe
exchange couplings as a function of Fe-Fe separation for
NdFe$_{12}$ and NdFe$_{12}$N. (b) The same data set is plotted
on a plane spanned by $J_{ij}$(NdFe$_{12}$) and $J_{ij}$(NdFe$_{12}$N),
with $J_{ij}$ shared among NdFe$_{12}$ and NdFe$_{12}$N as the 
spatially equivalent bonds.
}
\end{figure}
\subsection{Derivation of the leading-order parameters}
As shown in Fig.~\ref{fig::Harashima_lattice},
the material NdFe$_{12}$N has a ThMn$_{12}$ crystal structure (space group $I4/mmm$)
with Nd occupying the body-centered site of a tetragonal unit
which incorporate 2 formula units. 
Around each Nd site, there are 4 Fe($8i$) atoms which
surround along the $a$-axis and $b$-axis. Four Fe($8j$) atoms make a square right in the middle of Nd sites
along the $c$-axis and nitrogen goes in the center of this square. Fe($8f$) sites make a tetragonal box
with the center occupied by Nd.
Taking the ThMn$_{12}$ crystal structure as the starting point,
the crystal structure is optimized from first principles. 
Then the crystal field parameters on Nd$^{3+}$ are calculated on the basis of open-core description
for the $4f$-shell~\cite{harashima}.
Up to here the calculations are done with the {\it ab initio}
electronic-structure computational code package, QMAS~\cite{qmas}.
The parameters for the exchange-coupling
are calculated
with the {\it ab initio}
electronic-structure calculation code,
Machikaneyama (AkaiKKR)~\cite{akai-KKR},
using the Korringa-Kohn-Rostoker Green's function method,
following the prescription proposed by Liechtenstein {\it et al.}~\cite{lichtenstein_1987}.
Re-writing the term in Eq.~(\ref{eq::3d}) as
\(
J_{ij}^{\rm TT}{\mathbf S}_{i}\cdot
{\mathbf S}_{j} \equiv 
(S_{i}J_{ij}^{\rm TT}S_{j})
{\mathbf e}_{i}\cdot
{\mathbf e}_{j}
\)
with the vector
${\mathbf e}_{i}$ denoting the direction of the magnetic moment
on site $i$,
calculated exchange couplings $(S_{i}J^{\rm TT}_{ij}S_{j})$
are summarized in Fig.~\ref{fig::Akai_Jij}~(a)
as a function of inter-site distance. As is the case
for the champion magnet compound Nd$_{2}$Fe$_{14}$B~\cite{rmp_1991},
we find that the dominant Fe-Fe exchange couplings come from Fe pairs
whose interatomic distances almost coincide with that in ${\alpha}$-Fe.
In our model we incorporate those dominant
Fe-Fe exchange couplings, that can be classified into 8 classes on the realistic lattice
for NdFe$_{12}$(N) to make a leading-order description. Comparing the calculated exchange couplings
on the same bonds located on the equivalent positions in the unit cell between NdFe$_{12}$
and NdFe$_{12}$N as shown in Fig.~\ref{fig::Akai_Jij}~(b), nitrogenation-enhanced exchange couplings
are identified. Interestingly, we observe
those bonds are located along a kagom\'{e}-lattice network
spanned by Fe($8f$) and Fe($8j$), reminiscent of one-generation-back permanent-magnet materials family
represented by SmCo$_{5}$ with the strong magnetic anisotropy. The other leading-order
term in the overall Hamiltonian written as Eq.~(\ref{eq::realistic_two_sublattice_model})
is $4f$ single-ion MAE $D_{m}^{\rm R}\left(J_{m}^{z}\right)^2$, which also
shows up in the energy scale of $O(10)$~[meV].
Re-writing the term in Eq.~(\ref{eq::4f}) as
\begin{equation}
D_{m}^{\rm R}\left(J_{m}^{z}\right)^2\equiv
K_{m}^{\rm R}\left(
{\mathbf e}_{m}
\cdot
{\mathbf n}
\right)^2,
\label{eq::defining_K_Nd}
\end{equation}
with ${\mathbf n}$ denoting the direction of the easy axis
and ${\mathbf e}$ the direction of the total magnetic moment ${\mathbf J}_{m}$
on the rare-earth site $m$,
a QMAS calculation gives 
$K^{\rm Nd}=5.8$ or $11.0~\mbox{[meV]}$ (easy-axis) for NdFe$_{12}$N
and 
$K^{\rm Nd}=-3.0$ or $-2.3~\mbox{[meV]}$ (easy-plane) for NdFe$_{12}$
depending on the calculation setups~\cite{harashima}.
As a first step, we choose 
$K^{\rm Nd}=8.1~\mbox{[meV]}$ for NdFe$_{12}$N and 
$=-2.8~\mbox{[meV]}$ for NdFe$_{12}$ within the range.

\begin{table}
\begin{center}
\caption{\label{table::J_RT}
Calculated $4f$-$3d$ exchange couplings $J^{\rm RT}_{mi}$ in meV.
The notation follows Eq.~(\ref{eq::J_RT}).
The site index
$m$ denotes
the Nd($2a$) sublattice.
}
\begin{tabular}{ccc}\hline
$i$ & NdFe$_{12}$ & NdFe$_{12}$N \\ \hline
Fe($8i$) & 1.58 & 1.90 \\ \hline
Fe($8j$) & 1.54 & 1.00  \\ \hline
Fe($8f$) & 1.93 & 2.36   \\ \hline
\end{tabular}
\end{center}
\end{table}

\subsection{Sub-leading parameters} We define $J^{\rm RT}$ and $D^{\rm T}$
in the overall Hamiltonian written as Eq.~(\ref{eq::realistic_two_sublattice_model}).
Calculated antiferromagnetic
exchange coupling \if0 within Liechtenstein's approach\fi
between $5d$-bands of Nd and $3d$-bands of Fe
are taken as they are to be the $5d$-mediated $4f$-$3d$ coupling,
assuming that intra-atomic direct exchange coupling is big enough to let the localized $4f$-moment
and $5d$-polarization work as a unified body. The results are summarized in
Table~\ref{table::J_RT}.
The Fe-originated MAE
in Eq.~(\ref{eq::3d}),
which we will denote analogously to Nd-originated one in Eq.~(\ref{eq::defining_K_Nd})
as follows,
\begin{equation}
D_{i}^{\rm T}\left(S_{i}^{z}\right)^2\equiv
K_{i}^{\rm T}\left(
{\mathbf e}_{i}
\cdot
{\mathbf n}
\right)^2
\label{eq::defining_K_Fe}
\end{equation}
comes in the order $O(0.1)$~[meV] as
referring to the past experimental measurements~\cite{nikitin_1998}
done for YFe$_{11}$Ti.
In the present modeling we just incorporate $K^{\rm Fe}=0.1~\mbox{[meV]}$ uniformly on
all Fe sites as a phenomenological setting
which should be sufficient
to describe the magnetic properties around the room temperatures and slightly higher.

Higher-order terms in addition to $K^{\rm Nd}$ in Eq.~(\ref{eq::defining_K_Nd}) can be suspected
in principle
for which we have also calculated and saw that
they are an order of magnitude smaller than the leading-order ones. Within the present scope to
pick up the leading-order behavior of finite-temperature observables, we have
dropped the higher-order contributions to the single-ion magnetic anisotropy of Nd.

\subsection{Methods} 
Having defined the realistic model basically
from first principles for NdFe$_{12}$(N), 
the temperature dependence of the magnetic properties
are calculated using the classical Monte Carlo method
with the Metropolis local updates.
We do one of the most plain local updates,
that is, picking up lattice sites stochastically
and proposing uniformly on the spherical spin space
up to the stochastic decision referring to the energy difference
as calculated by the spin model Hamiltonian defined as Eq.~(\ref{eq::realistic_two_sublattice_model}).
Effective one lattice sweep, {\it i.e.} with the stochastic choice of the lattice site
to be updated done $N_{\rm site}$ times, is counted as one Monte Carlo step.
Our input parameters
are simply set to the $T=0$ {\it ab initio} values.
For typical runs in the present work,
$10^5$ Monte Carlo steps (MCS) with each step consisting of a lattice sweep
were sufficient to reach the thermal equilibrium state and subsequent
$10^6$ MCS are used for the numerical measurement of observables.
The tetragonal unit of the body-centered-network of Nd is counted as
a system-size unit
in our calculation, thus a calculation with $L\times L\times L$ system
actually contains $26L^3\equiv N_{\rm site}$ spins with the tetragonal unit
having two formula units. Majority of the data are taken with $L=4$
which is sufficient as seen below with Fig.~\ref{fig::calculated_Ha_of_T}~(a)
while
the data points on
the high-temperature side
took system sizes up to $L=10$, with the anisotropy field getting smaller
and the calculated system getting closer to the Curie temperature where
the finite system size can become an issue due to the diverging
correlation length.

\section{Results}

With the prescription for the construction of the
realistic-lattice spin model, calculated temperature dependence
of magnetization and anisotropy field are shown in comparison with
the experimental data~\cite{hirayama_2014}. Model parameter dependence
of the overall calculated temperature dependence is inspected
to draw our main conclusion: $4f$-$3d$ exchange couplings $J^{\rm RT}_{mi}$
in Eq.~(\ref{eq::J_RT}) dominates the observables in the operation temperature range
{\it i.e.} around the room temperature or higher.

\subsection{Magnetization}
\begin{figure}
\begin{center}
\begin{tabular}{l}
(a) \\
\scalebox{0.7}{
\includegraphics{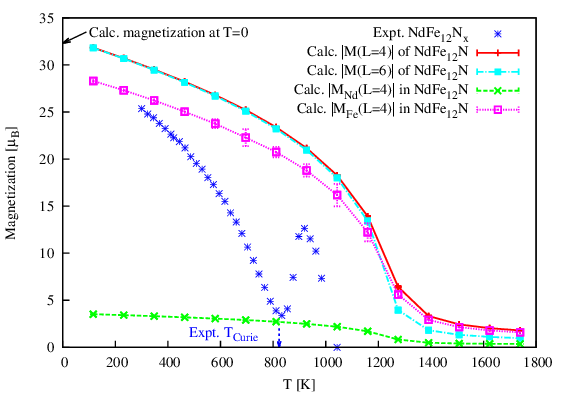}}
\\
(b) \\
\scalebox{0.7}{
\includegraphics{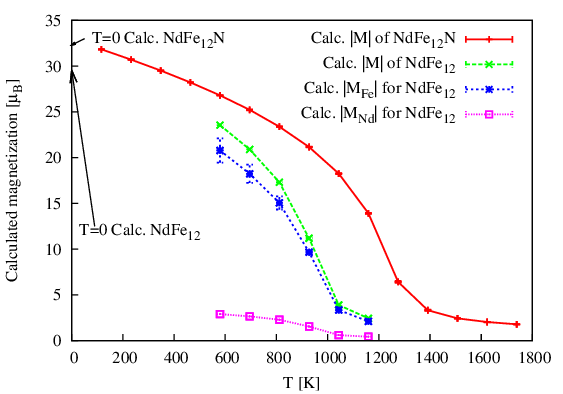}}
\end{tabular}
\end{center}
\caption{\label{fig::Curie temperarure} (a)
Calculated temperature dependence of magnetization for NdFe$_{12}$N
compared to the latest experimental data~\cite{hirayama_2014}.
(b) Comparison of the
calculated temperature dependence of magnetization
between NdFe$_{12}$ and NdFe$_{12}$N within $L=4$.}
\end{figure}

Calculated temperature dependence of magnetization $M(T)\equiv \sqrt{M_x^2+M_y^2+M_z^2}$ 
for the bulk, the Nd-sublattice, and the Fe-sublattice of NdFe$_{12}$N
is shown in Fig.~\ref{fig::Curie temperarure}~(a). Calculated Curie temperature
falls in the same range as the experimentally claimed one. Considering the
possible presence of $\alpha$-Fe-originated noise in the experimental data
and neglect of all next-nearest-neighbor exchange interactions in our model
which should have lowered the computational magnetic-ordering energy scale,
the agreement is satisfactory. Also we see that the system-size dependence
is negligible in our focus temperature range, which up to $T=500~\mbox{[K]}$ at most.
Comparing the calculated temperature dependence of magnetization between
NdFe$_{12}$ and NdFe$_{12}$N as shown in Fig.~\ref{fig::Curie temperarure}~(b),
we see that the experimentally observed~\cite{hirayama_2014}
nitrogenation-triggered enhancement in the Curie temperature
by $\sim 200$~[K] is well reproduced.

We note that Nd in NdFe$_{12}$ has the easy-plane magnetic anisotropy~\cite{hirayama_2014,harashima}
which would compete against the easy-axis anisotropy from at least a part of the Fe sublattices.
Here we just track the origin of
the difference in the Curie temperature between NdFe$_{12}$
and NdFe$_{12}$N to the nitrogenation-enhanced exchange couplings
as demonstrated in Fig.~\ref{fig::Akai_Jij}.
Further studies on the outcome of the competing anisotropies in NdFe$_{12}$
is separated
for future work.

\subsection{Anisotropy field}
\begin{figure}
\begin{center}
\begin{tabular}{l}
(a) \\
\scalebox{0.7}{
\includegraphics{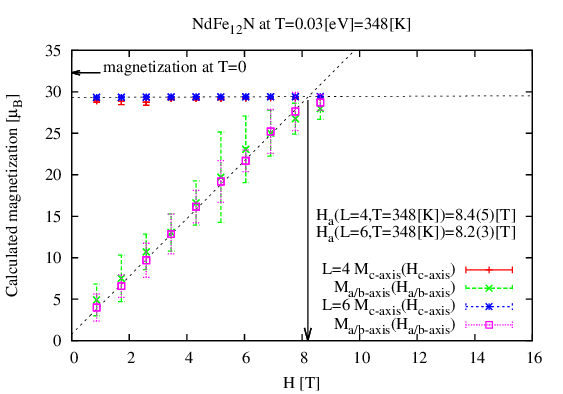}}\\
(b) \\
\scalebox{0.7}{
\includegraphics{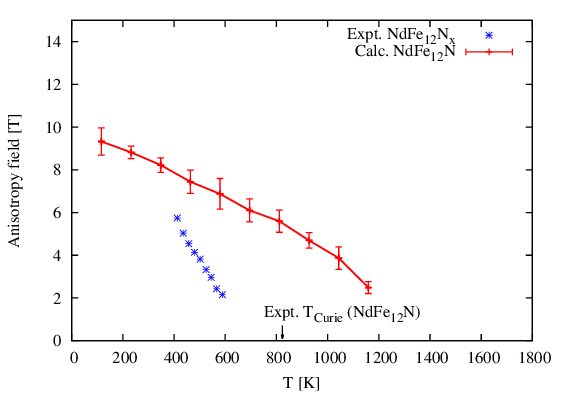}}
\end{tabular}
\end{center}
\caption{\label{fig::calculated_Ha_of_T} 
(a) Numerical determination
of the anisotropy field via the calculated magnetization curves
for NdFe$_{12}$N at $T=348$~[K] and
(b) thus calculated temperature dependence
of the anisotropy field for NdFe$_{12}$N.}
\end{figure}

Calculated magnetization curves with the externally
applied magnetic field parallel and perpendicular to the
easy axis in NdFe$_{12}$N (which is $c$-axis) at $T=348$~[K]
is shown in Fig.~\ref{fig::calculated_Ha_of_T}~(a).
The anisotropy field $H_a$ is identified
as a crossing point of the two curves with a linear fit.
Our numerical measurement just follows the experimental way to determine $H_a$.
Thus determined
anisotropy field $H_a(L,T)$
with the $L$-dependence being saturated out within the statistical error bars
is plotted as a function of temperature in Fig.~\ref{fig::calculated_Ha_of_T}~(b)
in comparison to the recent experimental data~\cite{hirayama_2014}.
In the operation temperature range,
a leading-order numerical
control seems to be achieved.

\subsection{Roles of each parameter}

\begin{figure}
\begin{center}
\begin{tabular}{l}
(a) \\
\scalebox{0.7}{
\includegraphics{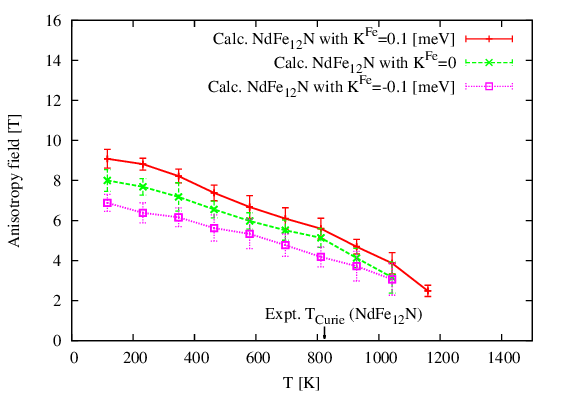}}\\
(b) \\
\scalebox{0.7}{
\includegraphics{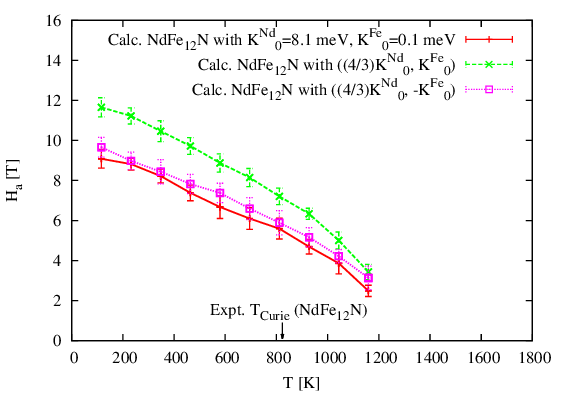}}
\end{tabular}
\end{center}
\caption{\label{fig::K_Nd_K_T_dependence} $(K^{\rm Nd},K^{\rm Fe})$-trends
of the calculated temperature dependence of the
anisotropy field for NdFe$_{12}$N. (a) Trends of $H_a(T)$
with respect to $K^{\rm Fe}=\pm 0.1$~[meV]. Data with $K^{\rm Fe}=0$ is also included
as a reference. (b) $K^{\rm Nd}$ is enhanced by a fraction of 1/3
in observing the trends of $H_a(T)$
with respect to $K^{\rm Fe}=\pm 0.1$~[meV].}
\end{figure}

We inspect which parameter dominates which part of the
temperature dependence of the observables comparing with the latest
experimental data. One of the most important practical utilities of permanent magnets
lies in the coercivity that is roughly proportional to the anisotropy field of the parent
material or the main phase, and the main issue here is
to figure out how to sustain the anisotropy field at high temperatures.
Calculated MAE for Nd, $K^{\rm Nd}$, and the
input parameters for $K^{\rm Fe}$ obviously
influence the temperature dependence of the anisotropy field. Flipping
the sign of $K^{\rm Fe}$ for the Fe sublattice to have an
easy-plane anisotropy, the overall reduction
of the anisotropy field by $2\times 12\times |K^{\rm Fe}|/K^{\rm Nd}\sim 30\%$
at the lowest temperature range
is observed as shown in Fig.~\ref{fig::K_Nd_K_T_dependence}~(a).
An interesting observation is made by enhancing $K^{\rm Nd}$ by 1/3 and flipping
the sign of $K^{\rm Fe}$ at the same time, where the overall anisotropy field
at the low-temperature side almost collapses onto the original temperature dependence
as shown in Fig.~\ref{fig::K_Nd_K_T_dependence}~(b).
The contribution of $K^{\rm Fe}$ is found to be quantitatively important
in the bulk MAE at finite temperatures where an increase in the anisotropy field only
by 10\% can lead to a practical breakthrough in designing LRE-based magnets~\cite{hono_2012}.
{\it Ab initio} determination of the exact nature of Fe-sublattice-dependent MAE is under way.

\begin{figure}
\begin{center}
\begin{tabular}{l}
(a) \\
\scalebox{0.7}{
\includegraphics{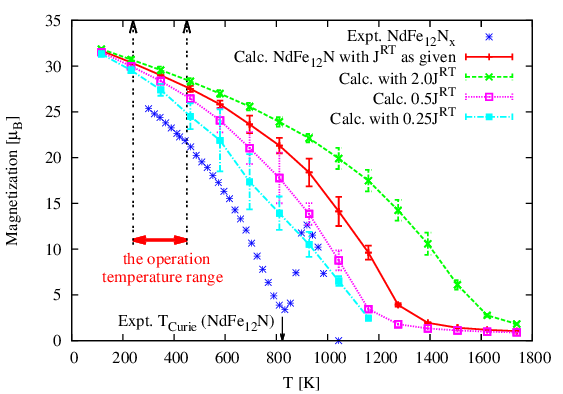}}
\\
(b) \\
\scalebox{0.7}{
\includegraphics{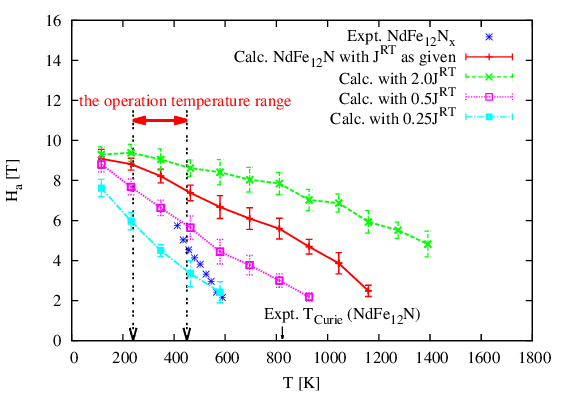}}
\end{tabular}
\end{center}
\caption{\label{fig::J_RT_dependence} $J^{\rm RT}$-dependence
of the calculated temperature dependence of (a)
magnetization and (b) anisotropy field for NdFe$_{12}$N. Given values from
the {\it ab initio} electronic-structure calculations are tabulated in Table~\ref{table::J_RT}.}
\end{figure}
Numerically manipulating the $5d$-mediated $4f$-$3d$
indirect
exchange coupling, $J^{\rm RT}$,
we observe how the calculated temperature dependence of
the magnetization and the anisotropy field in NdFe$_{12}$N is affected
as shown in Figs.~\ref{fig::J_RT_dependence}~(a)~and~(b), respectively.
The anisotropy field is more significantly affected by $J^{\rm RT}$
than the magnetization is 
especially around $T\sim 300$~[K].
This is
a numerical
demonstration that
a small enhancement in $J^{\rm RT}$,
which may be realized by
the conduction-band engineering in $4f$-$3d$ intermetallics,
can already lead to a considerable improvement.

\section{Discussions}

Having identified the quantitative relevance of
$J_{mi}^{\rm RT}$ written as
Eq.~(\ref{eq::J_RT}) for the magnetic properties of
NdFe$_{12}$N
in the operation temperature range,
we further
pin-down the role of
$J_{mi}^{\rm RT}$
in determining the realistic magnetic anisotropy energy scale at finite temperatures,
through a close comparison between calculated results and the experimental data.

In Fig.~\ref{fig::J_RT_dependence},
it is seen that
the experimentally observed slope and the downward convexity of
the temperature dependence of the anisotropy field for NdFe$_{12}$N
is closest to the calculated $H_a(T)$ with $\alpha_{\rm RT}=0.25$.
Also the experimental anisotropy field near the liquid-nitrogen temperature
seems to come close to 14~[T]~\cite{hirayama_confidential} 
which suggests possible
systematic underestimate in our calculated magnetic anisotropy
which falls below $10~\mbox{[T]}$ in the limit $T\rightarrow 0$.
The {\it ab initio} estimation for the crystal-field coefficients for the estimation
of the uni-axial magnetic anisotropy energy
indeed involves a certain uncertainty~\cite{harashima}.

\begin{figure}
\begin{center}
\scalebox{0.7}{
\includegraphics{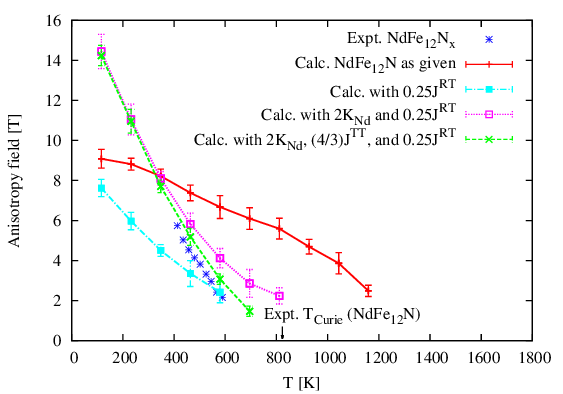}}
\end{center}
\caption{\label{fig::addendum}
For NdFe$_{12}$N, a data set which describes
the experimental data most quantitatively comes
with $\alpha_{\rm RT}=0.25$.}
\end{figure}

An improved data collapse between theory and experiment in the lowest temperature range
of $H_{a}(T)$ can be observed by manually
setting
$K^{\rm Nd}=2K^{\rm Nd}_{0}$.
As shown in Fig.~\ref{fig::addendum}, the slope of the temperature dependence
up to the room-temperature range is well reproduced.
The upper shift of the calculated $H_a(T)$ on the high-temperature
side of the operation temperature range can be adjusted
by a manual scaling of $J^{\rm TT}$ with an overall factor of 4/3
to match the Curie temperature. This can be considered as an effective
renormalization of the $3d$-$3d$ exchange couplings imposed by the discarded longer-range part in the exchange couplings.

Thus an inspection of the experimental temperature dependence of magnetization
and anisotropy field leads to a set of model parameters
for a quantitative description of the experimental
finite-temperature data. Based on the obvious relevance of $J^{\rm TT}$
for the Curie temperature and $K^{\rm R}$ for $H_a(T=0)$,
it is seen that $J^{\rm RT}$ determines the slope of $H_a(T)$ near $T\stackrel{>}{\sim}0$.

In such parameter set, the $4f$-$3d$ exchange couplings
come close to a quarter of the tabulated numbers which were $1\sim 2$~[meV] as seen
in Table~\ref{table::J_RT}.
In the language of the two-sublattice model~\cite{radwanski} where an isolated $4f$-electron magnetic moment 
is put into the sea of $3d$-electron magnetization via a molecular field
${\mathbf H}_{\rm m}$,
the molecular field can be written
in terms of our lattice model language
as follows.
\[
{\cal H}_{\rm m}=\alpha_{\rm RT}2J^{\rm RT}{\mathbf S}\times z_{\rm Nd}
\]
where $z_{\rm Nd}$ is the coordination number around the Nd magnetic moment which is in the present case
$z_{\rm Nd}=20$ within the nearest neighbor. Plugging in the realistic numbers
$\alpha_{\rm RT}\sim 0.25$, $J^{\rm RT}\sim 2$~[meV], $S\sim 2~\mbox[\mu_{\rm B}]$ the magnitude of the exchange field
is found to be $H_{\rm m}\sim 40~\mbox{\rm [meV]}=700~\mbox{\rm [T]}$ which gives the same order
as was found in the experimental analyses $H_{\rm m}\sim 450-600~\mbox{\rm [T]}$ by neutrons~\cite{loewenhaupt_1991}.
$\alpha_{\rm RT}=1$ corresponds to $H_{\rm m}$ beyond $1000~\mbox{\rm [T]}$ thus is out of the realistic scale.
The smallness
of the realistic number in the factor ${\alpha}_{\rm RT}$
presumably reflects the indirect nature
of the $4f$-$3d$ exchange.

%\section{Discussions}
%\label{sec::disc}

%
%SmCo$_5$ where
%both Sm and Co have uni-axial magnetic anisotropy

%We note in passing that

\section{Conclusions and outlook}
{\it Ab initio} modeling for the $4f$ and $3d$ magnetism coupled by $5d$-electrons
on the basis of the realistic spin-lattice model
for the rare-earth permanent magnet materials NdFe$_{12}$N and NdFe$_{12}$
quantitatively captures the realistic energy scales in the leading order
in the operating temperature range, $200~\mbox{[K]}\stackrel{<}{\sim}
T\stackrel{<}{\sim} 500~\mbox{[K]}$.
Experimentally observed
magnetic ordering and magnetic anisotropy energy scales are put under numerical
control
and we have shown
that enhancing the $4f$-$3d$ indirect exchange coupling 
would work most effectively
to realize the magnetic properties
of more practical use
in the operation temperature range.
Establishing a quantitative description starting
with the simplified model would pave the way to more realistic
simulations to explicitly incorporate the strongly-correlated nature
of $4f$-electrons embedded in the conduction-electron sea of $5d$-electrons
to comprehensively describe the finite-temperature physics
starting from the range $T\stackrel{>}{\sim}1~\mbox{[K]}$ all the way
to the Curie temperature close to $1000~\mbox{[K]}$.

\begin{acknowledgments}
Helpful comments given by A.~Sakuma, Y.~Hirayama,
and S.~Hirosawa, and collaborative works in related projects
with R.~Banerjee
and J.~B.~Staunton
are gratefully acknowledged.
This work is supported by the Elements Strategy Initiative Project under the auspice of MEXT, 
MEXT HPCI Strategic Programs for Innovative Research (SPIRE) and Computational Materials
Science Initiative (CMSI).
\end{acknowledgments}

\end{document}